\documentclass{epl2} 
\usepackage{amsmath}
\usepackage{bm}
\usepackage{graphicx}
\usepackage{subfigure}
\usepackage{amsmath}
\usepackage{amssymb}
\usepackage{mathrsfs}
\usepackage{braket}

\title{Fluctuations of topological charges in two-dimensional classical Heisenberg model} 
\shorttitle{Fluctuations of topological charges in 2D classical Heisenberg model}

\author{Shan-Chang Tang\inst{1,2}  
 \and   Yu Shi\footnote{yushi@fudan.edu.cn}\inst{3,1}}
\institute{ \inst{1} Department of Physics \&  State Key Laboratory of Surface Physics, Fudan University, Shanghai  200433, China\\
     \inst{2} Shanghai High School, Shanghai 200231, China\\   
         \inst{3} Wilczek Quantum Center, Shanghai Institute for Advanced Studies,  Shanghai 201315, China     
  }

\abstract{Binding and unbinding of vortices drives  Kosterlitz-Thouless phase transition in  two-dimensional XY model. Here we investigate whether similar mechanism works in   two-dimensional Heisenberg model, by using the fluctuation of skyrmion number inside a loop to characterize  the nature of  binding versus unbinding of defects. Through Monte Carlo simulations, we find that the fluctuation is proportional to the perimeter of the loop at low temperatures while it is proportional to the area of the loop at high temperatures, implying binding of the defects at low  temperatures and unbinding at high temperatures.
}

\begin{document}

\maketitle 

In 1966, Mermin and  Wagner presented the theorem later bearing their names,  that ``a one- or two-dimensional isotropic spin-S Heisenberg model with finite-range exchange interaction can be neither ferromagnetic nor antiferromagnetic'' at any nonzero temperature \cite{WagnerMermin-1966}. However, they also commented that  only spontaneous magnetization of sublattice magnetization had been ruled out, while the possibility of other kinds of phase transitions was not.  In 1973,  Kosterlitz-Thouless (KT) transition,  driven by the binding and unbinding of vortices, was  discovered  in two-dimensional XY model \cite{KosterlitzThouless-1972,KosterlitzThouless-1973,Kosterlitz-1974}.

Many investigations have been  made on whether there is a phase transition at a finite temperature in   two-dimensional Heisenberg model, and this remains controversial \cite{Kenna-2006}. Its  possible existence was proposed  according to the  high-temperature expansion \cite{StanleyKaplan-1966}. But its absence was predicted in various other approaches, including renormalization group \cite{Polyakov-1975,Zinn-JustinBrezin-1976B,Zinn-JustinBrezin-1976L,PelcovitsNelson-1977,TobochnikShenker-1980,BiferalePetronzio-1989}, perturbation theory \cite{Takahashi-1987,AllesBuonanno-1997,AllesCella-1999} and finite-size scaling \cite{Tomita-2014}, nevertheless these methods have also been used by other authors in claiming the opposite \cite{PatrascioiuSeiler-1998,Seiler-2003,
KapikranianBerche-2007}.
A recent analysis of the spin-spin correlation functions based on the tensor network method  found  supports both  for its presence and for its  absence~\cite{Schomoll-2021}.   In Monte Carlo simulations, the  magnetization versus temperature  relation is consistent with the absence  \cite{BlumeVineyard-1970,Betsuyaku-1978,
Klenin-1979,BrownCiftan-1993}. However, a sharp peak in the susceptibility suggests the possibility of the presence, and  it remains unclear whether the peak becomes singular at the thermodynamic limit \cite{Klenin-1979,BrownCiftan-1993}. The magnetization and susceptibility as functions of temperature  both shift  to lower temperatures  as the size increases \cite{BlumeVineyard-1970,Betsuyaku-1978}, and it is  unclear whether they peak at a nonzero temperature in the thermodynamic limit. The specific heat \cite{Colot-1983,BrownCiftan-1993} as a function of temperature exhibits a smoothed  peak that  is  scale-independent,  very similar to that in 2D XY model  \cite{KawabataBinder-1977}.  It was conjectured that topological defects   may  drive a   phase transition in 2D Heisenberg model \cite{BrownCiftan-1993}. Vortex-like defects were spotted at low temperatures, but it was not sure whether they can lead to a phase transition \cite{Klenin-1979,KawabataBishop-1980}. 

However, skyrmions cannot be  stable topological defects in 2D Heisenberg model,  as the dimension of the spin is larger than the spatial dimension \cite{Toulouse-1976}.  There is escape to the third dimension, and  it can be continuously deformed to a trivial state. 

In this Letter, we perform a Monte Carlo simulation  of 2D classical Heisenberg model, and investigate the temperature dependence of the nature of binding of topological defects, by using the fluctuation of skyrmion number as the characteristic quantity. It will be shown that   there are skyrmion fragments  with nonzero fluctuations of skyrmion numbers.   At  low temperatures,   skyrmion fragments with opposite charges  are paired with total charges zero. At high temperatures they are free.  Therefore, although skyrmion fragments are not topologically protected, 
they can drive a transition in the 2D Heisenberg system, in a way similar to the KT transition driven by vortices in the 2D XY system.

\begin{figure}[ht!]
  \centering
  \includegraphics[scale=0.19]{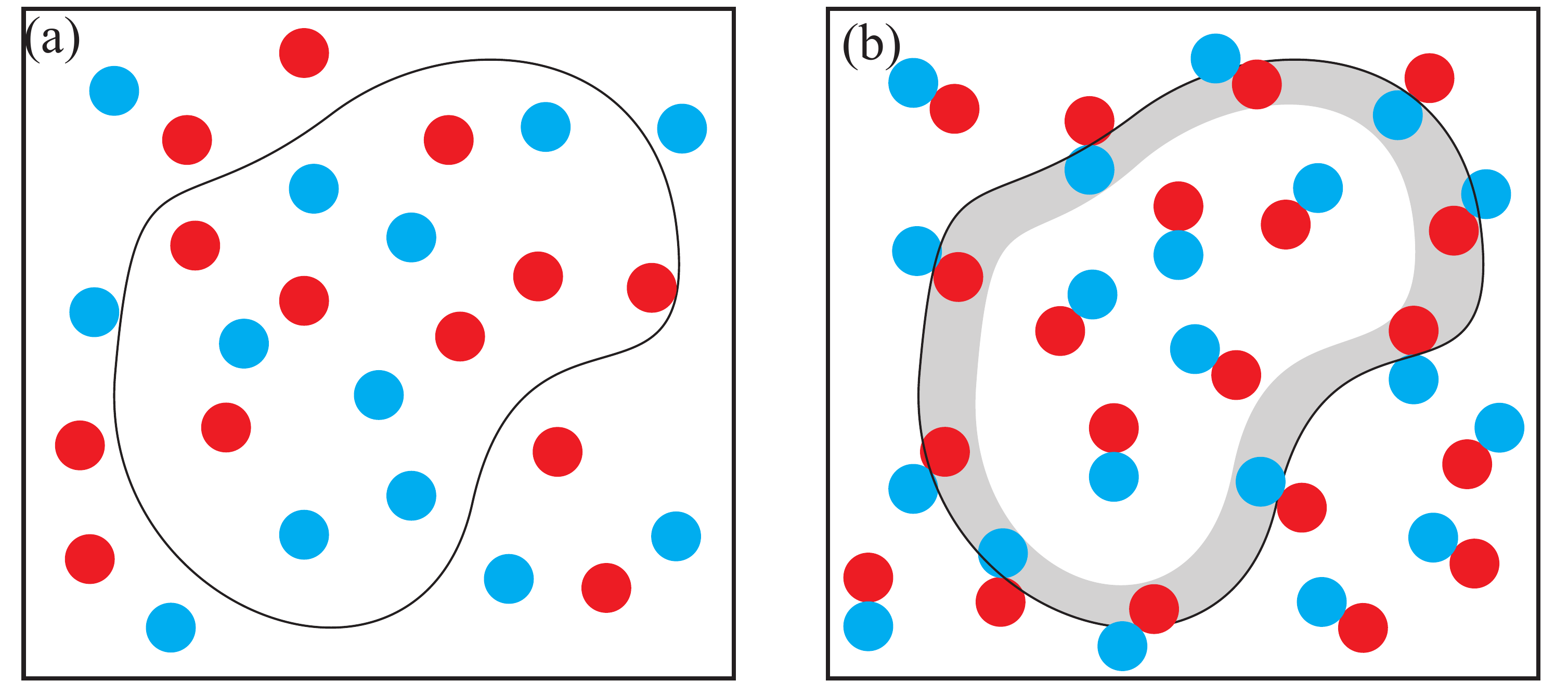}\\
  \caption{(a) The distribution of the topological defects at a high temperature. (b) The distribution of the topological defects at a low temperature. Red circles represent positive topological charges while blue ones are negative. The black line represents an arbitrary loop in the two-dimensional space.}\label{fig:long_range_order}
\end{figure}

We first introduce the method by using the 2D  XY model as the instructive example. The Hamiltonian  is
\begin{equation}
  \mathcal{H}^{{\rm XY} }=-J\sum_{\langle ij\rangle}\bm{S}_i\cdot\bm{S}_j,
\end{equation}
where $J>0$ is the ferromagnetic exchange constant, $\bm{S}_i=(S_{ix},S_{iy})$ represents the two-component classical spins on  site $i$, with  $\bm{S}_i^2=1$,  $\langle ij\rangle$ represents the nearest neighbours. It is well known that there exists a KT transition, driven by the binding and unbinding of vortices and antivortices. A vortex and an antivortex,  with topological charges $\pm 1$, attract each other, with  interaction proportional to the logarithm of their distance. They  are bound together at low temperatures, and become unbound and  move freely at high temperatures, as shown in Fig.~\ref{fig:long_range_order}.
 
Consider an arbitrary loop $\mathcal{C}$, with perimeter $l$ and area $\mathcal{S}$. Inside the loop,
the total topological charge is $Q=N_+-N_-$, where   $N_+$ and $N_-$ are the numbers of the vortices and the antivortices respectively.
Because of the randomness,  $\langle N_+\rangle=\langle N_-\rangle$ at high temperatures. Consequently,
$\langle Q\rangle_{\text{high T}}=\langle N_+ - N_-\rangle=0.$
The fluctuation of the topological charge $Q$ is
\begin{equation}
  \chi^{{\rm XY}}=\langle(Q-\langle Q\rangle)^2\rangle.
  \end{equation}
As KT transition is driven by the binding and unbinding of topological defects, the fluctuation of topological charge can be used to characterize KT transition~\cite{KosterlitzThouless-1972,KosterlitzThouless-1973}, 
just like the fluctuation of magnetization can be used to characterize the phase transition of ferromagnetism. 

At high temperatures, it can be found that
\begin{equation}
  \chi_{\text{high T}}^{{\rm XY}}=\langle(Q-\langle Q\rangle)^2\rangle_{\text{high T}}=(\langle Q^2\rangle-\langle Q\rangle^2)_{\text{high T}}\propto \mathcal{S},
\end{equation}
where  $\mathcal{S}$ is the area of the loop, for the following reason.
The fluctuation of  $Q$  is $
  \langle Q^2\rangle-\langle Q\rangle^2
  =(\langle N_+^2\rangle-\langle N_+\rangle^2)+(\langle N_-^2\rangle -\langle N_-\rangle^2)
  -2\langle N_+N_-\rangle+2\langle N_+\rangle\langle N_-\rangle=(\langle N_+^2\rangle-\langle N_+\rangle^2)+(\langle N_-^2\rangle -\langle N_-\rangle^2)$, where
 $\langle N_\pm\rangle=\frac{\sum_{rs}N_{\pm r} e^{\beta\mu_\pm N_{\pm r}-\beta E_s}}{\sum_{rs}e^{\beta\mu_\pm N_{\pm r}-\beta E_s}}$  is the average number in the grand canonical ensemble theory,  $\mu_\pm$ is the chemical potential for two kinds of topological defects, and it is assumed that the two kinds of topological defects are uncorrelated. As $k_BT \left(\frac{\partial\langle N_\pm\rangle}{\partial{\mu}_\pm}\right)_{T,V}
 =\langle N_\pm^2\rangle-\langle N_\pm\rangle^2$. In terms of the  unit volume $v_\pm=\frac{V}{\langle N_\pm\rangle}$, $
  \langle N_\pm^2\rangle-\langle N_\pm\rangle^2=-\frac{k_BTV}{v_\pm^2}\left(\frac{\partial v_\pm}{\partial \mu_\pm}\right)_T.$  With
  $G_\pm=\mu_\pm \langle N_\pm\rangle$ , $dG_\pm=-SdT+Vdp+\mu_\pm d\langle N_\pm\rangle$, we have $d\mu_\pm=-s_\pm dT+v_\pm dp$, where $s_\pm\equiv \frac{S}{\langle N_\pm\rangle}$ is the unit entropy.
  As a result, we have  $\left(\frac{\partial\mu_\pm}{\partial v_\pm}\right)_T=\left(\frac{\partial\mu_\pm}{\partial p}\right)_T\left(\frac{\partial p}{\partial v_\pm}\right)_T=v_\pm\left(\frac{\partial p}{\partial v_\pm}\right)_T$, therefore   $
  \langle N_\pm^2\rangle-\langle N_\pm\rangle^2=-\frac{k_BTV}{v_\pm^3}\left(\frac{\partial v_\pm}{\partial p}\right)_T=\frac{k_BT\kappa_T}{v_\pm^2}V
  \equiv C_\pm V,$
where all factors in $C_\pm$ are all intensive quantities, while  $V$  becomes the area  $\mathcal{S}$ in two dimensions. Consequently, the fluctuation of the topological charge is
$\chi^{{\rm XY}}=(C_+ +C_-)\mathcal{S}\propto\mathcal{S}.$

At low temperatures, the topological defects are bounded as  pairs (Fig.~\ref{fig:long_range_order}(b)). If the two defects of a  pair  are  both inside the loop, their  charges cancel each other, thus do not  contribute to the total topological charge inside the loop. Only the  inner member of each pair crossing the boundary contributes  to  the total topological charge.

Because of randomness, at low temperatures, the average topological charge is also zero
\begin{equation}
  \langle Q\rangle_{\text{low T}}=\langle N_+ - N_-\rangle=0.
\end{equation}

However, since only the topological defects around the boundary from the inner side contribute, as shown in the grey belt in Fig.\ref{fig:long_range_order}(b),  the fluctuation of the charge is proportional to the length $l$ of the loop,
\begin{equation}
  \chi_{\text{low T}}^{{\rm XY}}=\langle(Q-\langle Q\rangle)^2\rangle_{\text{low T}}=(\langle Q^2\rangle-\langle Q\rangle^2)_{\text{low T}}\propto l,
\end{equation}
for the following reason.  As can be seen from Fig.~\ref{fig:long_range_order}(b), $Q=Q_{\text{in}}+Q_{\text{cross}}=Q_{\text{cross}}$, where $Q_{\text{in}}$ represents the topological charge of the defects that totally reside in the loop and $Q_{\text{cross}}$ represents the charge inside the gray belt on the graph. The area of this belt is $l\times D$, with $D$ being the average size of the defects. As a result, the fluctuation in this case is
 $ \chi_{\text{low T}}^{{\rm XY}}=\langle Q^2\rangle_{\text{low T}}-\langle Q\rangle^2_{\text{low T}}=\langle Q_{\text{cross}}^2\rangle_{\text{low T}}-\langle Q_{\text{cross}}\rangle^2_{\text{low T}}
  \propto\mathcal{\mathcal{S}_\text{belt}}\propto l.$

Therefore, if  the fluctuation of the topological charge inside a loop scales with the perimeter of the loop, then it is indicated that  the vortices and the antivortices are bound.

Now we consider the two-dimensional classical  isotropic Heisenberg model, in which each spin possesses three components,  $\bm{S}=(S_{ix},S_{iy},S_{iz})$. Its Hamiltonian is
\begin{equation}
  \mathcal{H}^H=-J\sum_{\langle ij\rangle}\bm{S}_i\cdot\bm{S}_j.
\end{equation}

Considering  the possibility of  topological defects in this  model \cite{Klenin-1979,KawabataBishop-1980}, we conjecture that there may be skyrmion  fragments, with the   total topological charge, or the total number,  inside an arbitrary loop being 
\begin{equation} Q_{\text{continuous}}=\iint_{(x,y)\in\Omega}\bm{S}\cdot\left(\frac{\partial\bm{S}}{\partial x}\times\frac{\partial\bm{S}}{\partial y}\right)dxdy,
\end{equation}
where $\Omega$ is the two-dimensional space inside the loop.

\begin{figure}[ht!]
  \centering
  \includegraphics[scale=1]{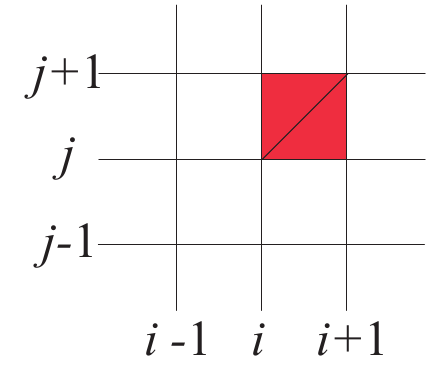}\\
  \caption{A sketch of the two-dimensional square lattice.}\label{fig:square_lattice}
\end{figure}

On a discrete square lattice, the definition of the  topological charge is discretized \cite{BergLuscher-1981}. As shown in Fig.\ref{fig:square_lattice}, there is a  three-component normalized classical spin $\bm{S}_{ij}$ on each site $(i,j)$, where $i$ and $j$ represent the row   and the column, respectively.  The on-site topological charge density $Q_{ij}$ can be defined  as the sum of the two solid angles of the spins on the vertices of the two red triangles,
\begin{equation}
    Q_{ij}=\alpha\left(\bm{S}_{ij},\bm{S}_{i+1,j},\bm{S}_{i+1,j+1}\right)+\alpha\left(\bm{S}_{ij},\bm{S}_{i+1,j+1},\bm{S}_{i,j+1}\right)\label{equ:topo_charge_density}
\end{equation}
In the simulation, we choose a square loop with  edge length $d$, and the total topological charge inside the loop is
\begin{equation}
  Q=\sum_{1\leqslant i\leqslant d,1\leqslant j\leqslant d}Q_{ij}.
\end{equation}
In case  there is a complete skyrmion inside the loop, $Q$  is equal to $4\pi$. This confirms that  the definition \eqref{equ:topo_charge_density} is reasonable.

Following the reasoning similar to that for the XY model, we can make the following hypotheses for  the two-dimensional Heisenberg model. (1) Area law: At  high temperatures,  the defects  can move freely, the fluctuation of the total topological charge inside an arbitrary loop is proportional to the area surrounded  by the loop,  $
        \chi_{\text{high T}}^{H}=\langle(Q-\langle Q\rangle)^2\rangle_{\text{high T}}=(\langle Q^2\rangle-\langle Q\rangle^2)_{\text{high T}}\propto \mathcal{S}.$
(2) Perimeter law: At low temperatures, if the defects of  opposite topological charges are bound, the fluctuation of the total topological charge inside an arbitrary loop is proportional to the length of the loop.
$\chi_{\text{low T}}^{H}=\langle(Q-\langle Q\rangle)^2\rangle_{\text{low T}}=(\langle Q^2\rangle-\langle Q\rangle^2)_{\text{low T}}\propto l.$
We can then calculate these physical quantities in the Monte Carlo simulations.

We perform the Monte Carlo simulation using the standard Metropolis algorithm on an $L\times L$ square lattice with the periodic boundary condition ($L=32,64,128,256,512$). We start with a random spin configuration at a temperature $k_BT=2J$, where $k_B$ is the Boltzmann constant. Then we gradually lower the temperature until it approaches zero.    For each value of the temperature, there are $10^5$   computational steps  per spin,   the ensemble average of the physical quantities is made over $10^4$ samples. The number of the intermediate sweeps between two samples is $30$. Here the intermediate sweeps, which   are included to decrease the correlation between the chosen configurations, refer to the sweeps between two configurations that are used to calculate the thermal quantities.

Several thermodynamic quantities are numerically  calculated in the simulations.
Most of these  results are similar to previous works, indicating that our simulation method is plausible.

\begin{figure}[ht!]
  \centering
  \includegraphics[scale=0.15]{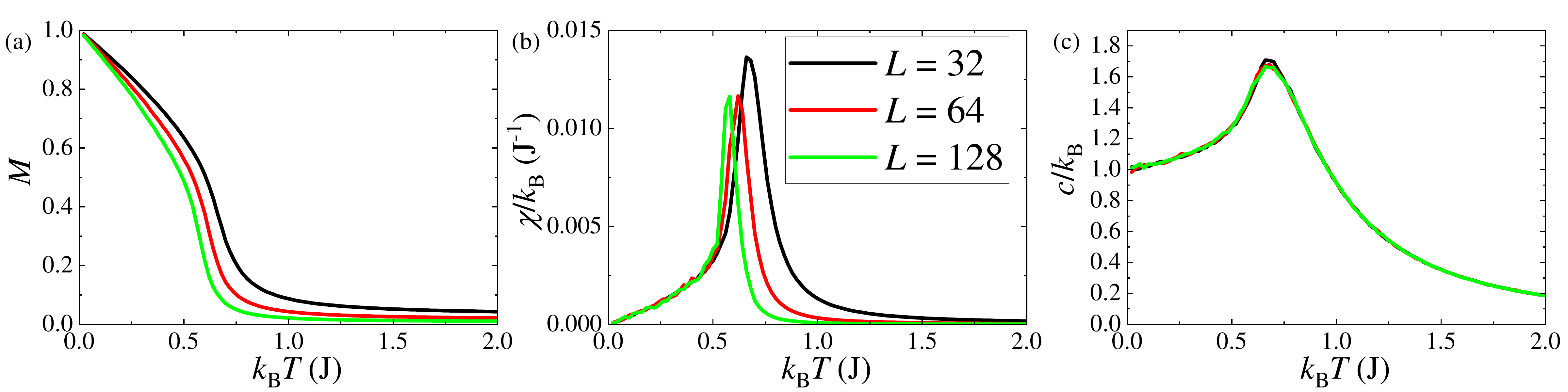}\\
  \caption{(a) The magnetization. (b) The susceptibility. (c) The specific heat capacity as function of temperature. The plots with different colors represent different lattice sizes: $32\times32$, $64\times64$, $128\times128$.}\label{fig:thermodynamic_quantities}
\end{figure}

First, the  magnetization  $M=\langle\sqrt{\bm{m}\cdot\bm{m} }\rangle$, where
   $\bm{m}=\frac{1}{L^2}\sum_i \bm{S}_i$
  is the average magnetic moment for a certain spin configuration, the susceptibility $    \chi=\frac{\langle\bm{m}\cdot\bm{m}\rangle-M^2}{T}$, and the specific heat capacity $
    c=\frac{1}{L^2}\frac{\langle E^2\rangle-\langle E\rangle^2}{k_BT^2}$,
  where $E$ is the total energy of the system.
We have obtained these three thermodynamic quantities as functions of  temperatures, for different lattice sizes (Fig.~\ref{fig:thermodynamic_quantities}).

With the increase of the lattice size, the magnetization as a function of the temperature shifts towards lower temperature (Fig.~\ref{fig:thermodynamic_quantities}(a)), in  consistency with  \cite{BlumeVineyard-1970,Betsuyaku-1978}, the susceptibility, which displays a sharp peak,  also shifts towards lower temperature (Fig.~\ref{fig:thermodynamic_quantities}(b)), in  consistency with  \cite{Klenin-1979,BrownCiftan-1993}, the  specific heat capacity, exhibiting a smoothed peak, remains  independent of the lattice size, in consistency with \cite{Colot-1983,BrownCiftan-1993}.

Moreover, in our  simulation, the specific heat capacity approaches $k_B$ as the temperature approaches zero, in consistency with  the equipartition theorem, as  there are two degrees of freedom in the $O(3)$ Heisenberg model.

In one of the previous  simulations \cite{Colot-1983}, the specific heat capacity, which is called ``heat capacity'' there,  approaches $k_B$, in consistency with our result. In another   simulation  \cite{BrownCiftan-1993},  the limit of the specific heat is $2$, which seems to be with a different definition.

\begin{figure}[ht!]
  \centering
  \includegraphics[scale=0.4]{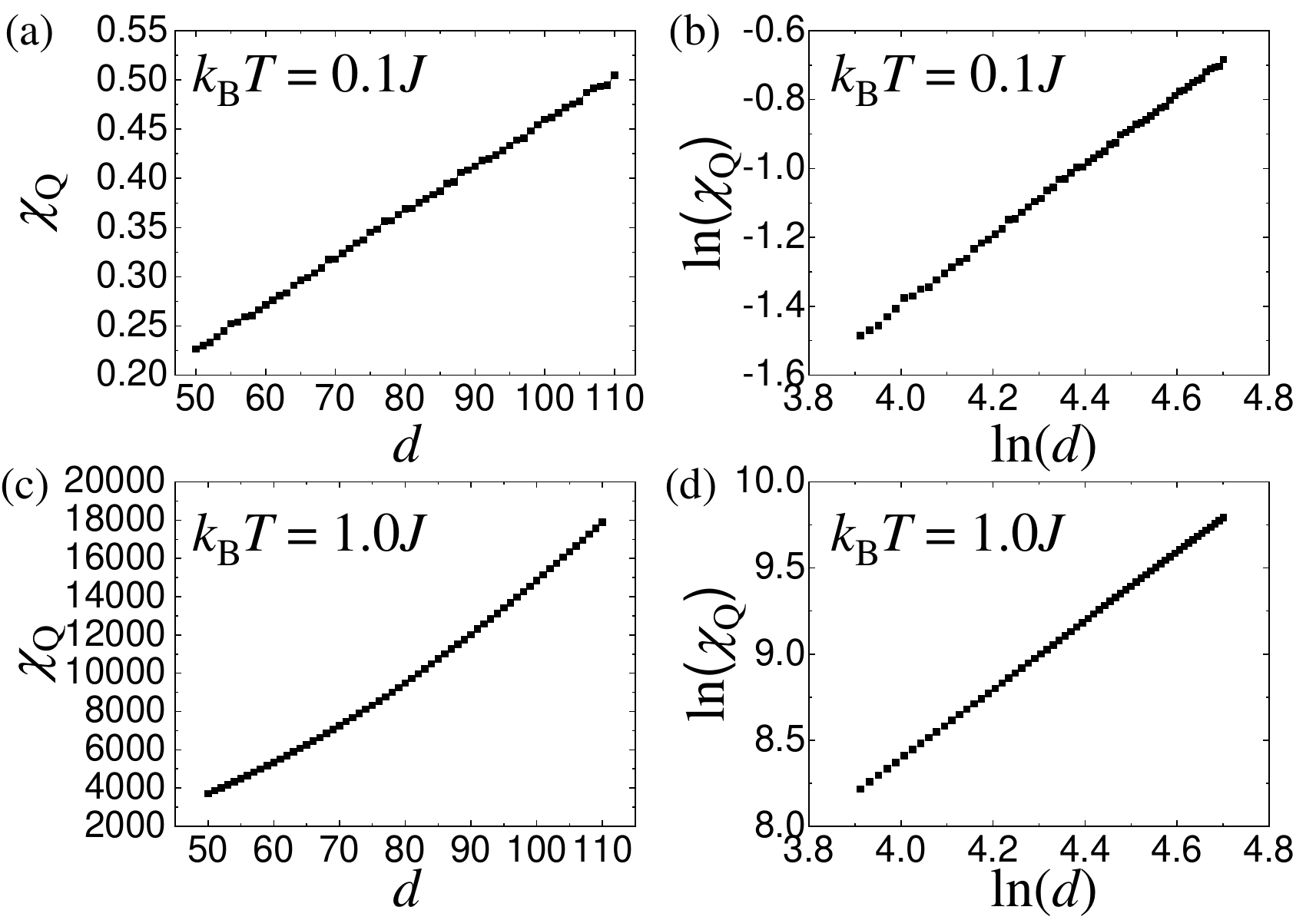}\\
  \caption{(a) The topological charge fluctuation versus the linear size $d$ of the loop at  $k_BT=0.1J$. (b) Plot of  (a) in terms of the logarithms of the fluctuation and  $d$. (c)The topological charge fluctuation versus the linear size $d$  at   $k_BT=1.0J$. (d) Plot of   (c) in terms of the logarithms of the fluctuation and  $d$. The lattice  size  is  $L=128$.  }\label{fig:chi_Q-d-relations}
\end{figure}

Now we come to the charge fluctuation.
For each value of the temperature, we   calculate  the charge fluctuation  $\chi^{H}$ for various  values of  $d$. For $L=128$, the results for   $k_BT=0.1J$ and $k_BT=1.0J$ are shown in Fig.~\ref{fig:chi_Q-d-relations}(a) and (c),  respectively. It is clear that the dependence on    $d$ is different for high and low temperatures.

Furthermore, for various values of the temperature, the relation between the logarithm of the fluctuation and  the logarithm of the loop size (Fig.~\ref{fig:chi_Q-d-relations}b and Fig.~\ref{fig:chi_Q-d-relations}d) is well fitted by a linear function,
\begin{equation}
  \ln(\chi^H)=k\cdot\ln(d),
\end{equation}
where the  $k \approx 1.01769$ for $k_BT=0.1J$ while it is $k \approx 2.00114$ for $k_BT=1.0J$. This implies that  at a lower temperature, $\chi^{H}_{\text{Low T}}\propto d^{1.01769}$, while at a higher temperature, $\chi^{H}_\text{High T}\propto d^{2.00114}$. That is, the fluctuation is approximately proportional to the loop size at a lower temperature while  approximately proportional to the square of the loop size  at a higher temperature.  Since the perimeter is proportional to the the loop size $l=4d$ and the area is the square of the loop size $\mathcal{S}=d^2$, the above result confirms our conjecture that the topological charge fluctuation satisfies the perimeter law at a lower temperature,  while it satisfies the  area law at a   higher temperature.

\begin{figure}[ht!]
  \centering
  \includegraphics[scale=0.5]{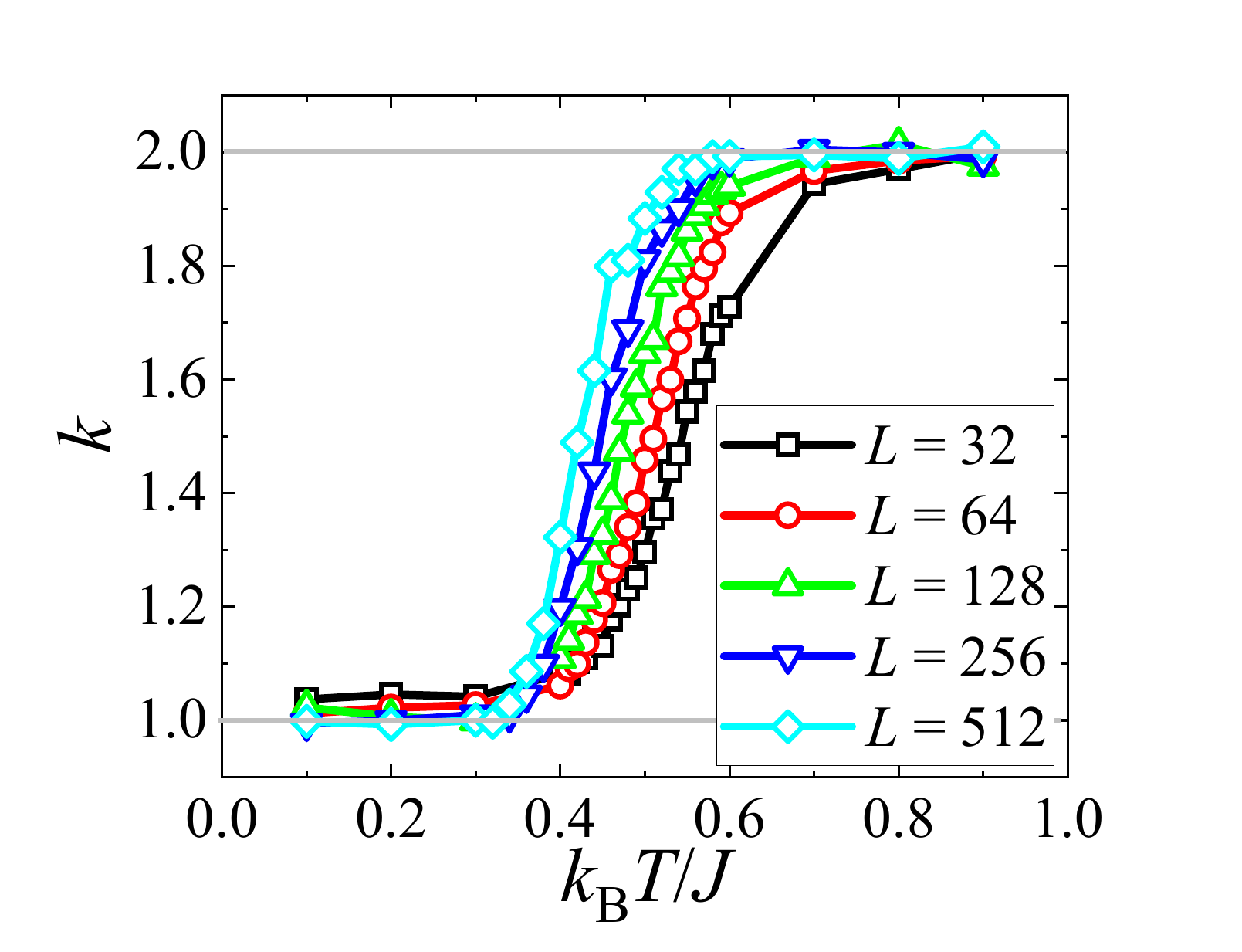}\\
  \caption{The slope $k$ of  the logarithm  of the fluctuation as a linear function of the logarithm of  $d$, as a function of  temperature for  different lattice sizes. The range of $d$ for $L=32$ is from 10 to 20; for $L=64$ is from 20 to 50; for $L=128$ is from 50 to 110; for $L=256$ is from 110 to 240; for $L=512$ is from 240 to 500.}\label{fig:k_T}
\end{figure}

For each of the five values of the lattice size, and for different temperatures, we obtain the slope  $k$ of the logarithm  of the fluctuation as a linear function of the logarithm of  $d$, thus we obtain $k$ as a function of temperature (Fig.~\ref{fig:k_T}).

The result clearly shows that the charge fluctuation satisfies the perimeter law  at low temperatures,  while it  satisfies   area law at high temperatures, as   conjectured above, moreover, there is a sharp  transition between these two topological phases and it becomes sharper   as the lattice becomes larger .

\begin{figure}[ht!]
  \centering
  \includegraphics[scale=0.6]{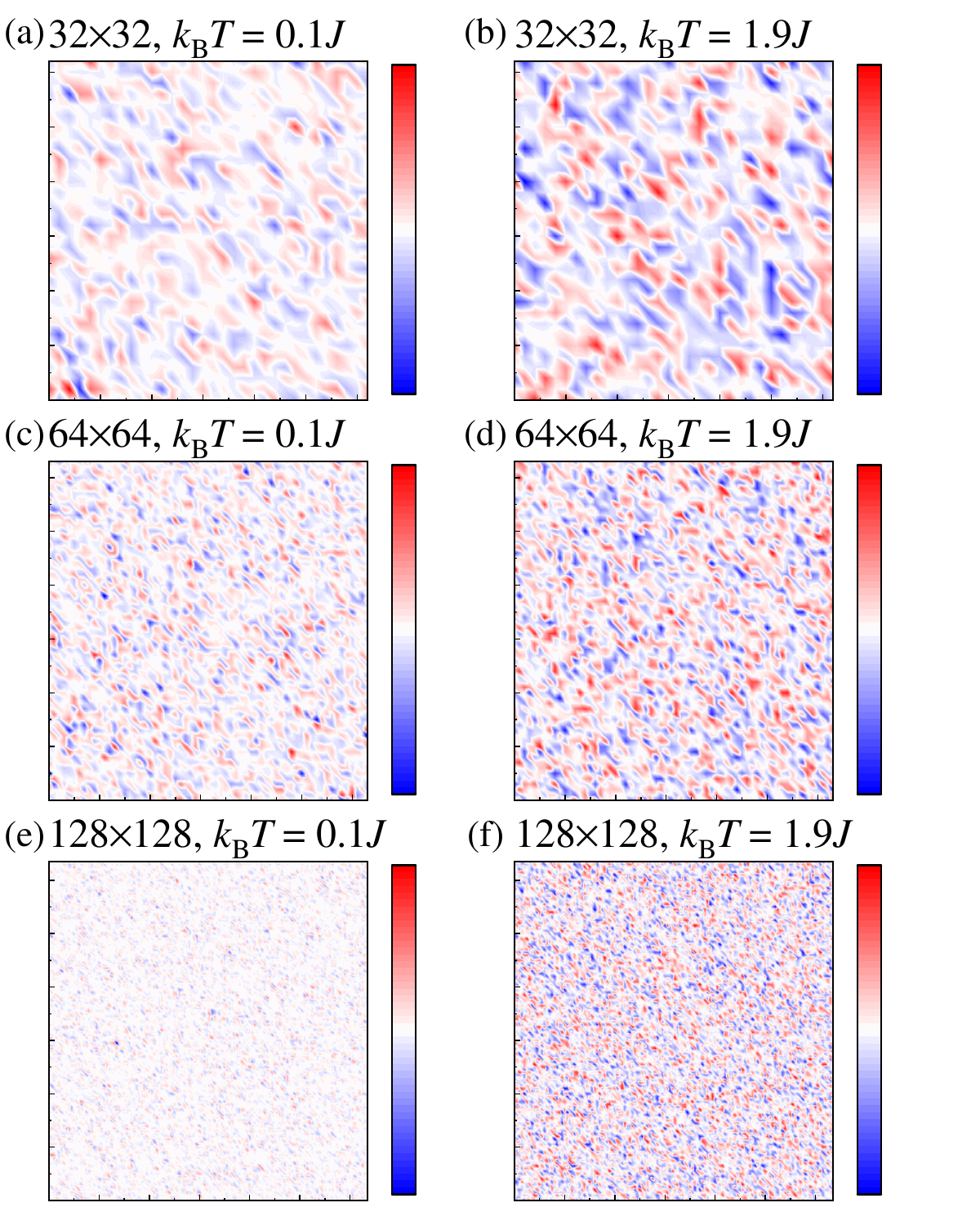}\\
  \caption{Distribution  of skyrmion density for various lattice sizes at various  temperatures.  A red point represents $Q_{ij}>0$ and a blue point represents $Q_{ij}<0$. (a)$L=32,k_BT=0.1J$. (b)$L=32,k_BT=1.9J$. (c)$L=64,k_BT=0.1J$. (d)$L=64,k_BT=1.9J$. (e)$L=128,k_BT=0.1J$. (f)$L=128,k_BT=1.9J$. }\label{fig:skyrmion_density}
\end{figure}

We have also calculated the distribution of skyrmion density $Q_{ij}$ for various lattice sizes at various  temperatures (Fig.~\ref{fig:skyrmion_density}).  In the simulation, we  note that indeed there are no complete skyrmions, rather, there are only fragments of the skyrmions, with nonzero topological charges. 

It is quite difficult to distinguish different topological phases from the patterns of  these skyrmion fragments. In fact, our method provides a  very useful  quantitative way to describe well  the binding and unbinding of the defects.

To summarize, we have performed Monte Carlo simulations to study various properties of the classical isotropic Heisenberg model in two dimensions. The magnetization, the magnetic susceptibility and the specific heat capacity are calculated for  different temperatures and are consistent with previous results. A novel physical quantity, the fluctuation of the topological charge inside a loop, is defined and calculated for different temperatures. We find that it is proportional to the area of the loop at  high temperatures while proportional to the perimeter of the loop at  low temperatures. This discovery strongly indicates that at low temperatures, the defects with opposite topological charges are bound together into a  composite particle with topological charge zero,  while they are unbound and can move freely at high temperatures. The binding versus  unbinding of defects implies  a possible transition in the two-dimensional Heisenberg model,  similar to KT transition in the two-dimensional XY model. In our simulation, the topological defects are skyrmion fragments, rather than complete skyrmions, but they can still pair into neutral composites. Our work sheds new light on the phases in the two-dimensional Heisenberg model.  The methods we use may also be applied to other systems such as three-dimensional Heisenberg model.

We thank Ping Ao, Siu-Tat Chui and John M. Kosterlitz for useful discussions. This work was supported by National Science Foundation of China (Grant No. 12075059).

\end{document}